\documentclass[prc,twocolumn,showkeywords,showpreprintnumbers,floatfix,superscriptaddress]{revtex4}
\usepackage{amsmath,amssymb,graphicx,bm,color,ulem}
\newcommand{\rigid}{\text{rigid}}
\newcommand{\irrot}{\text{irrot}}
\newcommand{\Gogny}{\text{D1S}}
\newcommand{\vek}[1]{\bm{\mathrm{#1}}}
\newcommand{\frakI}{\mathfrak{I}}
\DeclareMathOperator{\arcsinh}{arcsinh}
\begin{document}

%%%%%%%%%%%%%%%%%%%%%%%%%%%%%%%%%%%%%%%%%%%%%%%%%%%%%%%%%%%%%%%%%%%%%%%%%%%%%%%
\title{Macroscopic manifestations of rotating triaxial superfluid nuclei}
%%%%%%%%%%%%%%%%%%%%%%%%%%%%%%%%%%%%%%%%%%%%%%%%%%%%%%%%%%%%%%%%%%%%%%%%%%%%%%%
\author{P. Schuck}
\email{schuck@ipno.in2p3.fr}
\affiliation{Institut de Physique Nucl{\'e}aire, CNRS-IN2P3,
  Univ. Paris-Sud, Universit\'e Paris-Saclay, F-91406 Orsay Cedex, France}
\affiliation{Univ. Grenoble Alpes, CNRS, LPMMC, F-38000 Grenoble, France}
\author{M. Urban}
\email{urban@ipno.in2p3.fr}
\affiliation{Institut de Physique Nucl{\'e}aire, CNRS-IN2P3,
  Univ. Paris-Sud, Universit\'e Paris-Saclay, F-91406 Orsay Cedex, France}
%%%%%%%%%%%%%%%%%%%%%%%%%%%%%%%%%%%%%%%%%%%%%%%%%%%%%%%%%%%%%%%%%%%%%%%%%%%%%%%
\begin{abstract}
  Recently, Allmond and Wood [Phys. Lett. B \textbf{767}, 226 (2017)]
  were able to extract the three moments of inertia $\frakI_k$ of a
  dozen of superfluid triaxial nuclei from experimental data. The
  observed dependence of the $\frakI_k$ on the deformation parameters
  is rather smooth. Here we show that these moments of inertia can be
  surprisingly well explained by a semiclassical cranked
  Hartree-Fock-Bogoliubov (HFB) calculation in which the velocity
  field is a simple superposition of rigid and irrotational flows.
\end{abstract}

\maketitle
%%%%%%%%%%%%%%%%%%%%%%%%%%%%%%%%%%%%%%%%%%%%%%%%%%%%%%%%%%%%%%%%%%%%%%%%%%%%%%%

%%%%%%%%%%%%%%%%%%%%%%%%%%%%%%%%%%%%%%%%%%%%%%%%%%%%%%%%%%%%%%%%%%%%%%%%%%%%%%%
\section{Introduction}
%%%%%%%%%%%%%%%%%%%%%%%%%%%%%%%%%%%%%%%%%%%%%%%%%%%%%%%%%%%%%%%%%%%%%%%%%%%%%%%
It is a well known fact that superfluidity has an important influence on
nuclear rotation. This for instance induces a strong reduction of the
moment of inertia of superfluid nuclei by factors two, three, or more.

\begin{figure*}
\includegraphics[width=5.6cm]{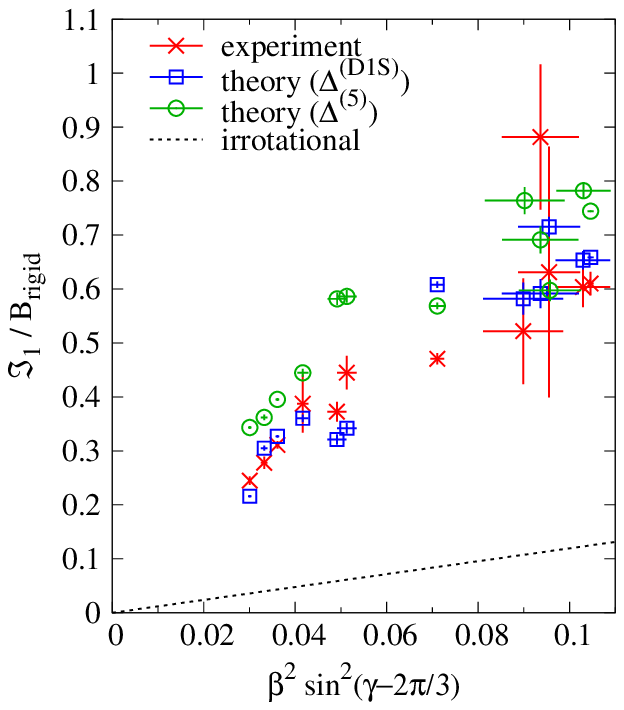}
\includegraphics[width=5.6cm]{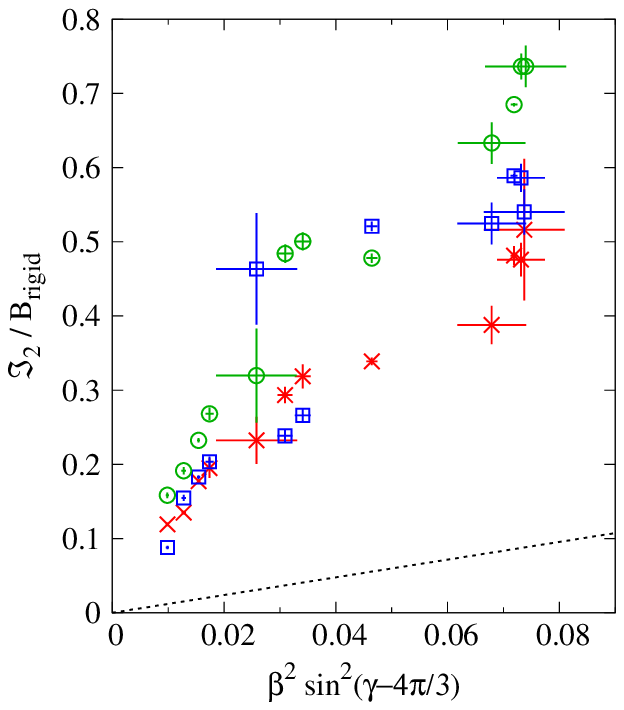}
\includegraphics[width=5.6cm]{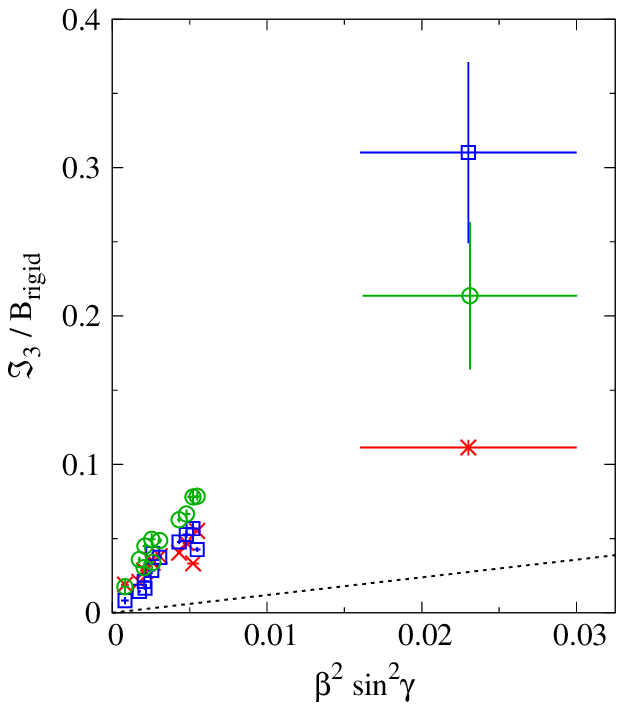}
\caption {\label{fig:J}Moments of inertia $\frakI_k$, plotted as a
  function of $\beta^2\sin^2(\gamma-\frac{2\pi}{3} k)$ which is
  proportional to $\frakI_{k,\irrot}$ (dashed line). The theoretical
  results were computed with the experimental deformation parameters
  $\beta$ and $\gamma$ \cite{Allmond2017} and the HFB pairing gaps
  $\Delta_{n,p}^{(\Gogny)}$ \cite{Hilaire-private} (blue boxes) and
  with the pairing gaps $\Delta_{n,p}^{(5)}$ from the 5-point formula
  (green circles). The error bars of the theoretical results include
  only the experimental uncertainties of $\beta$ and $\gamma$. The
  results are compared with the experimental moments of inertia of
  Ref.~\cite{Allmond2017} (red crosses).}
\end{figure*}

In the past this feature was revealed in the great majority of cases
for rotation of axially symmetric nuclei. Rotation of triaxially
deformed nuclei is much scarcer and less well born out (see, e.g.,
Ref.~\cite{Helgesson1989} for an early work and
Ref.~\cite{Washiyama2018} for one of the latest developments). Very
recently, Allmond and Wood \cite{Allmond2017} made a very nice
analysis of a dozen of the clearest triaxially deformed nuclei in
deducing experimentally their moments of inertia around the three
axes. In Fig.~\ref{fig:J},
%%%%%%%%%%%%%%%%%%%%%%%%%%%%%%%%%%%%%%%%%%%%%%%%%%%%%%%%%%%%%%%%%%%%%%%%%%%%%%%
%%%%%%%%%%%%%%%%%%%%%%%%%%%%%%%%%%%%%%%%%%%%%%%%%%%%%%%%%%%%%%%%%%%%%%%%%%%%%%%
we show a reproduction of their figure for the three moments of
inertia corresponding to the three axes. Besides very few exceptions
the experimental results (red crosses) lie with relatively little
scatter around a straight line. This new and surprising feature calls
for a simple explanation.

As early as in 1959, Migdal developed a statistical description of
rotating superfluid nuclei where he applied some sort of Strutinsky
smoothing of a superfluid contained in a deformed harmonic oscillator
potential while making also some estimates how things would change
with a hardwall box potential \cite{Migdal1959}. He was able to well
explain the general trend of the superfluid quadrupole moment of
inertia as a function of deformation and neutron number, see Fig. 1 in
\cite{Migdal1959}. In 1985, Durand, Schuck, and Kunz \cite{Durand1985}
translated Migdal's statistical approach into a semiclassical
transport model. The formulae for the moment of inertia stayed
unchanged, only it was then possible to also calculate the current
distributions. It was shown that the current distribution in rotating
superfluid nuclei evolves as a function of the gap value from rigid
rotation for small gaps to irrotational flow patterns for very large
gaps. Realistic values of the gap show intermediate features of the
flow.

From the late 90s, experimentalists achieved to produce atomic Bose-Einstein
condensates in traps. Anticipating that it would become possible to
trap also fermionic atoms and cool them down to superfluidity, Farine
et al. used the theory originally developed for nuclei to compute the
moment of inertia of atomic Fermi gases \cite{Farine2000}. In 2003, we
elaborated a more advanced semiclassical transport approach to
rotating superfluid fermionic atoms including the temperature
dependence \cite{Urban2003}. Indeed, a few years later, the experiment
of a rotating Fermi gas was realized by the Innsbruck group and the
reduction of the moment of inertia below the superfluid critical
temperature was observed \cite{Riedl2011}.

Concerning now the measurements of the moments of inertia in
triaxially deformed superfluid nuclei, we only had to reactivate our
past calculations and adopt them to the triaxial deformation. As we
will see, we get very good agreement with the experimental
values. Since these results come from a semiclassical approach where
shell effects are absent, the agreement between theory and experiment
reveals a {\it macroscopic} behavior of triaxially deformed
superfluid nuclei. This is, maybe, a somewhat surprising but nice
finding for such a subtle feature as is triaxial rotation. For
completeness let us repeat our analytic formulae which we will use in
order to explain the measurements of the moments of inertia as well as
those needed for the calculation of the flow patterns.

%%%%%%%%%%%%%%%%%%%%%%%%%%%%%%%%%%%%%%%%%%%%%%%%%%%%%%%%%%%%%%%%%%%%%%%%%%%%%%%
\section{Formalism}
%%%%%%%%%%%%%%%%%%%%%%%%%%%%%%%%%%%%%%%%%%%%%%%%%%%%%%%%%%%%%%%%%%%%%%%%%%%%%%%
As stated in the introduction, we repeat the harmonic oscillator model
of Migdal, generalized to triaxiality as in \cite{Farine2000}. The
starting point is a cranked HFB calculation to which a semiclassical
approximation is applied. The formulas (\ref{Inertia})-(\ref{current})
are given in our earlier publication \cite{Urban2003} and for
completeness repeated here. Let us start with the superfluid moment of
inertia, which is given by
\begin{equation}
  \frakI = \frakI_{\rigid}
  \bigg(1 - \frac{8\omega_x^2\omega_y^2G_+G_-}
       {(\omega_x^2 + \omega_y^2)(\omega_+^2G_+ + \omega_-^2G_-)}\bigg)\,,
    \label{Inertia}
\end{equation}
where $\omega_k$ are the frequencies of the triaxially deformed
harmonic oscillator potential and $\omega_{\pm} = \omega_y \pm
\omega_x$, $\frakI_{\rigid}$ is the corresponding rigid-body
moment of inertia, and the functions $G_{\pm}$ are given by
\begin{equation}
  G_{\pm} = G\Big(\frac{\hbar\omega_{\pm}}{2\Delta}\Big)\,,
\end{equation}
where $\Delta$ is the gap at equilibrium at the Fermi level and
\begin{equation}
  G(x) = \frac{\arcsinh(x)}{x\sqrt{1+x^2}}\,.
  \label{G}
\end{equation}
From Eq.~(\ref{Inertia}) one sees that in the limit of very strong
pairing, $\Delta\gg \hbar\omega_i$ (i.e., $G_\pm\to 1$), the moment of
inertia reduces to its irrotational value
\begin{equation}
  \frakI_{\irrot} =
  \bigg(\frac{\omega_x^2-\omega_y^2}{\omega_x^2+\omega_y^2}\bigg)^2
  \frakI_{\rigid}.\label{Iirrot}
\end{equation}
These formulas are arranged for a rotation around the $z$-axis, but
rotation about the other two axes is easily achieved in permuting the
axes.

The current corresponding to a rotation with angular velocity
$\vek{\Omega}= \Omega \vek{e}_z$ is given by
\begin{multline}
  \vek{j}(\vek{r}) = \rho(\vek{r})\vek{v}(\vek{r})\\
  = \Omega\rho(\vek{r}) \bigg (r_x\vek{e}_y - r_y\vek{e}_x -
  \frac{4G_+G_-(\omega_x^2r_x\vek{e}_y - \omega_y^2r_y\vek{e}_x}
         {\omega_+^2G_+ + \omega_-^2G_-}\bigg)\,.
  \label{current}
\end{multline}
Again, one sees that, with increasing gap, the velocity field changes
continuously from the rigid rotation
\begin{equation}
  \vek{v}_{\rigid} = \vek{\Omega}\times\vek{r}
  \label{eq:vrot}
\end{equation}
to the irrotational one
\begin{equation}
  \vek{v}_{\irrot} =
  \Omega\frac{\omega_y^2-\omega_x^2}{\omega_y^2+\omega_x^2}\vek{\nabla}xy\,.
  \label{eq:virrot}
\end{equation}

Equations (\ref{Inertia}) and (\ref{current}) can be summarized in the
compact form
\begin{gather}
  \frakI = (1-C)\, \frakI_\rigid + C\, \frakI_{\irrot}\,,\label{Ifinal}\\
  \vek{v} = (1-C)\, \vek{v}_\rigid + C\, \vek{v}_{\irrot}\,,\label{vfinal}\\
  C = \frac{2G_+G_- (\omega_x^2+\omega_y^2)}{\omega_+^2G_+ + \omega_-^2G_-}\,.
  \label{eq:C}
\end{gather}

%%%%%%%%%%%%%%%%%%%%%%%%%%%%%%%%%%%%%%%%%%%%%%%%%%%%%%%%%%%%%%%%%%%%%%%%%%%%%%%
\section{Results and discussion}
%%%%%%%%%%%%%%%%%%%%%%%%%%%%%%%%%%%%%%%%%%%%%%%%%%%%%%%%%%%%%%%%%%%%%%%%%%%%%%%
In order to produce numbers, we have to determine the parameters
entering Eq.~(\ref{Ifinal}). The rigid-body moment of inertia for
rotation about the $k$ axis ($k=1,2,3$ corresponding to $x,y,z$) is
approximated as (see, e.g., \cite{RingSchuck})
\begin{gather}
  \frakI_{k,\rigid} = B_{\rigid}\Big(1-\sqrt{\tfrac{5}{4\pi}}\beta
      \cos(\gamma-\tfrac{2\pi}{3} k)\Big)\,,\\
  B_{\rigid} = \tfrac{2}{5} m A R_0^2
      = 0.0138  A^{5/3}\, \hbar^2\, \text{MeV}^{-1}\,,
\end{gather}
with $m$ the nucleon mass, $R_0 = 1.2\,\text{fm}\, A^{1/3}$ the
nuclear radius, and $\beta$ and $\gamma$ the deformation parameters
(Hill-Wheeler coordinates). The oscillator frequencies $\omega_k$ are
inversely proportional to the radii, i.e., up to corrections of higher
order in $\beta$,
\begin{equation}
\hbar\omega_k = 41\,\text{MeV}\, A^{-1/3}
  \Big(1-\sqrt{\tfrac{5}{4\pi}}\beta\cos(\gamma-\tfrac{2\pi}{3} k)\Big)\,.
\end{equation}
To leading order in $\beta$, one thus obtains from Eq.~(\ref{Iirrot})
\begin{equation}
  \frakI_{k,\irrot} = \tfrac{15}{4\pi} B_{\rigid} \beta^2
    \sin^2(\gamma-\tfrac{2\pi}{3} k)\,.
\end{equation}

In this work, we will consider the 12 triaxial nuclei whose $\beta$
and $\gamma$ values are listed in Table 1 of Ref.~\cite{Allmond2017}
[we corrected a typo in that table: for $^{110}$Ru, the correct value
  is $\beta=0.310(11)$ as can be inferred from the value of the
  quadrupole moment $Q_0 = 3ZeR_0^2\beta/\sqrt{5\pi}$
  \cite{RingSchuck} given in the same table].

We furthermore need the pairing gaps. The fact that neutron and proton
gaps $\Delta_n$ and $\Delta_p$ are different can easily be accounted
for by replacing \cite{Migdal1959}
\begin{equation}
\frakI(\Delta) \to \frac{N}{A} \frakI(\Delta_n)+\frac{Z}{A}
\frakI(\Delta_p)\,.
\end{equation}
As explained in \cite{Farine2000}, what we need in our semiclassical
approximation are the average gaps on the Fermi surface. They can be
extracted, e.g., from HFB calculations with the D1S Gogny force
\cite{Hilaire-private} which describe the ground-state properties of
these nuclei (including deformations) very well. More precisely, we
denote by $\Delta_{n,p}^{(\Gogny)}$ the HFB gaps averaged with
$u^2v^2$ as explained in Ref.~\cite{Hilaire2002}. An alternative and
much simpler way to obtain values for the gaps is to compute them from
the experimental nuclear masses $M$ \cite{Wang2017} using the 5-point
formula \cite{Bender2000}
\begin{multline}
  \Delta_n^{(5)} = \tfrac{1}{8}[-M(N-2,Z)+4M(N-1,Z)-6M(N,Z)\\
    +4M(N+1,Z)-M(N+2,Z)]\,,
\end{multline}
and analogously for $\Delta_p^{(5)}$. In contrast to the simpler
3-point formula, this formula eliminates mean-field effects to a large
extent \cite{Bender2000}. As can be seen in Table~\ref{tab}, the
agreement with the HFB gaps is on the average not too bad, although
the $\Delta^{(5)}$ tend to be smaller than the $\Delta^{(\Gogny)}$.
%%%%%%%%%%%%%%%%%%%%%%%%%%%%%%%%%%%%%%%%%%%%%%%%%%%%%%%%%%%%%%%%%%%%%%%%%%%%%%%
\begin{table*}
  \caption{\label{tab} Values $\beta^2\sin^2(\gamma-\frac{2\pi}{3}k)$
    computed from $\beta$ and $\gamma$ given in \cite{Allmond2017},
    pairing gaps $\Delta_{n,p}^{(\Gogny)}$ computed with the D1S Gogny
    force \cite{Hilaire-private}, and $\Delta_{n,p}^{(5)}$ obtained
    from nuclear masses \cite{Wang2017} using the 5-point formula, for
    the 12 triaxial nuclei considered in Ref.~\cite{Allmond2017} and
    in the present paper.}
  \begin{ruledtabular}\begin{tabular}{llllllll}
  Nucleus&$\beta^2\sin^2(\gamma-2\pi/3)$&$\beta^2\sin^2(\gamma-4\pi/3)$&
    $\beta^2\sin^2\gamma$&$\Delta_n^{(\text{D1S})}$ (MeV)&
    $\Delta_p^{(\text{D1S})}$ (MeV)&$\Delta_n^{(5)}$ (MeV)&$\Delta_p^{(5)}$ (MeV)\\
  \hline
  $^{110}$Ru&0.096(7)&0.025(8)&0.023(8)&0.82&1.10&1.20&1.40\\
  $^{150}$Nd&0.0711(16)&0.0464(10)&0.00261(8)&1.08&0.90&1.05&1.19\\
  $^{156}$Gd&0.093(9)&0.068(7)&0.0021(3)&1.39&1.08&1.00&0.97\\
  $^{166}$Er&0.1046(7)&0.0719(6)&0.00306(14)&0.92&1.49&0.90&0.92\\
  $^{168}$Er&0.103(6)&0.073(5)&0.00254(24)&1.09&1.18&0.77&0.88\\
  $^{172}$Yb&0.090(9)&0.074(8)&0.0008(3)&1.25&1.17&0.73&0.89\\
  $^{182}$W &0.0513(22)&0.0341(15)&0.00175(11)&1.20&2.01&0.80&0.81\\
  $^{184}$W &0.0491(22)&0.0309(14)&0.00210(15)&1.23&2.08&0.75&0.85\\
  $^{186}$Os&0.0417(13)&0.0174(8)&0.0052(4)&1.30&1.01&0.91&1.01\\
  $^{188}$Os&0.0361(4)&0.0155(3)&0.00432(14)&1.28&1.02&0.96&1.00\\
  $^{190}$Os&0.0332(8)&0.0128(4)&0.00479(24)&1.31&1.00&0.97&1.06\\
  $^{192}$Os&0.0301(4)&0.0099(3)&0.00549(22)&1.44&1.45&0.91&1.11
  \end{tabular}\end{ruledtabular}
\end{table*}
%%%%%%%%%%%%%%%%%%%%%%%%%%%%%%%%%%%%%%%%%%%%%%%%%%%%%%%%%%%%%%%%%%%%%%%%%%%%%%%

In Fig.~\ref{fig:J} we show the resulting moments of inertia
$\frakI_k$ of the 12 triaxial nuclei considered in \cite{Allmond2017},
plotted as a function of the combination of deformation parameters
$\beta^2\sin^2(\gamma-\frac{2\pi}{3} k)$, which is proportional to
$\frakI_{k,\irrot}$, for the two choices of pairing gaps. To make the
identification between points in the figures and nuclei easier, we
have listed the values of $\beta^2\sin^2(\gamma-\frac{2\pi}{3} k)$ for each
nucleus in Table~\ref{tab}. For each of the three axes $k=1,2,3$, the
moments of inertia lie more or less on a smooth curve between
$\frakI_{k,\irrot}$ and $\frakI_{k,\rigid}$.

The overall agreement between theoretical (blue boxes and green
circles) and experimental (red crosses) moments of inertia is
surprisingly good. Of course there are some cases where it works less
well, as expected for a semiclassical theory which does not include
shell effects. In particular, there are some outliers, such as
$^{110}$Ru for which especially $\frakI_2$ and $\frakI_3$ are clearly
too large. In general, the moments of inertia computed with
$\Delta^{(5)}$ tend to be too large, which is related to the fact that
in most cases $\Delta^{(5)} < \Delta^{(\Gogny)}$. Maybe $\Delta^{(5)}$
is not always a good estimate for the average gap on the Fermi
surface.

\begin{figure*}
\includegraphics[width=5.6cm]{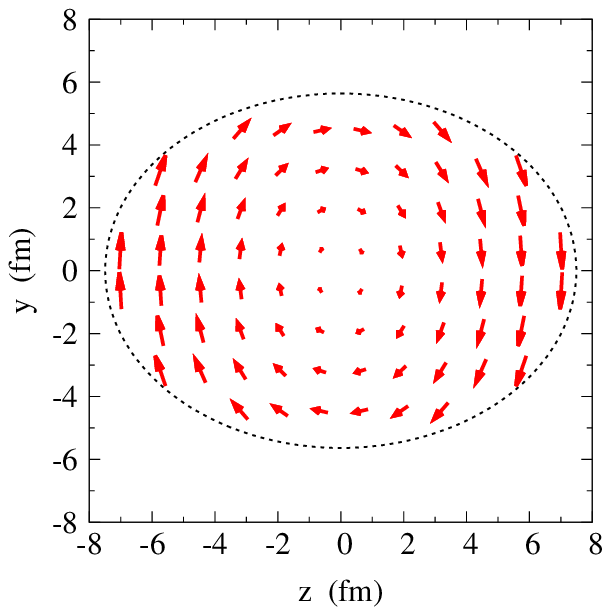}
\includegraphics[width=5.6cm]{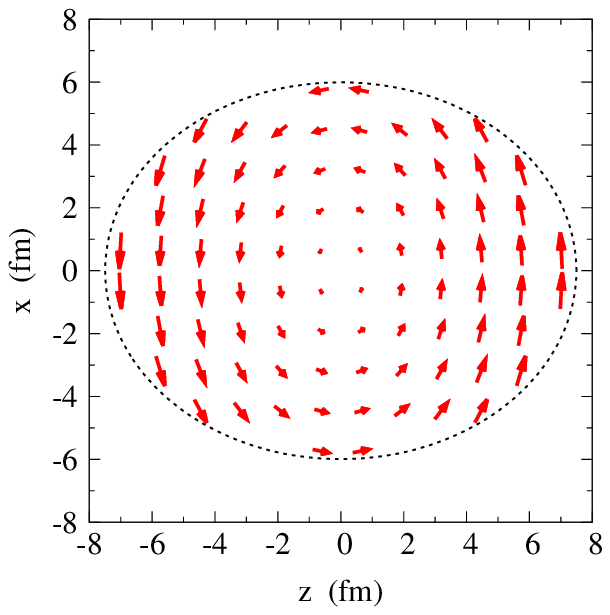}
\includegraphics[width=5.6cm]{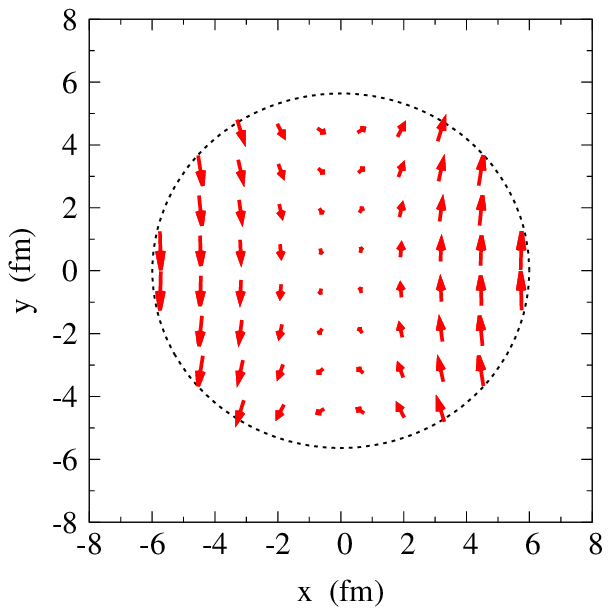}
\caption {\label{fig:v} Velocity fields in $^{150}$Nd for
  rotations about the three principal axes. For better visibility, in
  the first two panels, the angular momentum is larger by a factor of
  two than in the third panel.}
\end{figure*}

It is interesting to see where, for a given nucleus, the difference in
$\frakI_k/B_{\rigid}$ depending on the axis $k$ comes from. To answer
this question, one can look at the velocity fields. As an example, we
show in Fig.~\ref{fig:v}
%%%%%%%%%%%%%%%%%%%%%%%%%%%%%%%%%%%%%%%%%%%%%%%%%%%%%%%%%%%%%%%%%%%%%%%%%%%%%%%
%%%%%%%%%%%%%%%%%%%%%%%%%%%%%%%%%%%%%%%%%%%%%%%%%%%%%%%%%%%%%%%%%%%%%%%%%%%%%%%
the velocity fields computed for the nucleus $^{150}$Nd for rotations
about the three axes (the $\Delta_{n,p}^{(5)}$ pairing gaps were used
in this example). As one can see, the velocity field is neither that
of a rigid rotation nor purely irrotational. But in the case of the
rotation about the $x$ axis, the rotational component is clearly
larger than in the case of the rotation about the $z$ axis where the
velocity field is closer to the typical form of irrotational flow.
And since $\frakI_{\irrot}$ is very small if the nucleus is almost
symmetric with respect to the rotation axis, this explains why
$\frakI_3$ is so much smaller than $\frakI_1$. In the present example,
our calculation gives $\frakI_1/\frakI_3 \simeq 17$ which is close to
the ratio of the experimental moments of inertia $\frakI_1/\frakI_3
\simeq 14$.
%%%%%%%%%%%%%%%%%%%%%%%%%%%%%%%%%%%%%%%%%%%%%%%%%%%%%%%%%%%%%%%%%%%%%%%%%%%%%%%
\section{Conclusions}
%%%%%%%%%%%%%%%%%%%%%%%%%%%%%%%%%%%%%%%%%%%%%%%%%%%%%%%%%%%%%%%%%%%%%%%%%%%%%%%
In this work, we calculated the three moments of inertia of triaxial
superfluid nuclei as they were deduced experimentally in a recent
paper by Allmond and Wood \cite{Allmond2017}. We used for that a
semiclassical approach which we had developed earlier for rotating
superfluid atomic clouds \cite{Farine2000,Urban2003} and which is
actually based on a very early work of Migdal concerning rotating
superfluid axially symmetric nuclei \cite{Migdal1959}, see also
\cite{Durand1985}. In the case of cold atoms, a semiclassic approach
seems very well justified since the number of atoms can reach around a
million. In finite nuclei, expectation values of observables are often
overshadowed by strong shell fluctuations and a semiclassical approach
can only yield an average value. However, as the work by Allmond and
Woods shows \cite{Allmond2017}, apparently the moments of inertia
$\frakI_k$ exhibit a rather smooth behavior as a function of the
variable $\beta^2\sin^2(\gamma-\frac{2\pi}{3}k)$ (proportional to the
moment of inertia in the case of purely irrotational flow) where
$\beta$ and $\gamma$ are the Hill-Wheeler coordinates. So we
used our analytic formulas given in \cite{Farine2000} for the
calculation of the three moments of inertia for each one of the 12
nuclei considered in \cite{Allmond2017}. To our surprise, the
agreement with experiment can be judged as good to very good. Actually
the experimental data show rather little shell fluctuations what hints
to rotation of superfluid triaxial liquid drops. In this sense, a
semiclassical description may be valid for quantal objects as small as
nuclei. The moments of inertia lie half way in between rigid rotation
and irrotational flow. To reproduce this feature is not trivial at all
and confirms that triaxial nuclear rotation exhibits macroscopic
aspects.

One may wonder why this is so. Actually, out of all nuclei, the
triaxially deformed ones are closest to a nuclear liquid drop (absence
of shell effects). It is a well known fact that shell fluctuations
diminish with the number of broken symmetries. One can establish the
following hierarchy: spherical-normal fluid $\to$ spherical-superfluid
$\to$ axially deformed-superfluid $\to$ triaxially-deformed
superfluid. This hierarchy goes along with a more and more smooth
single-particle level density. Therefore, the semiclassical theory can
be expected to describe qualitatively and even semiquantitatively the
moments of inertia, and it is very exciting that this is so well
confirmed experimentally.

With our approach we were also able to calculate the flow
patterns. Not surprisingly, we see a mixture of irrotational and
rotational motion. Naturally the rotation around the axis with the
least deformation shows the most prononounced irrotational behavior
(and vice versa for the strongest deformation axis). It is very nice
that analytic formulas are able to catch essentially all the subtle
features of rotation of superfluid triaxial nuclei very well.

%%%%%%%%%%%%%%%%%%%%%%%%%%%%%%%%%%%%%%%%%%%%%%%%%%%%%%%%%%%%%%%%%%%%%%%%%%%%%%%
\begin{acknowledgments}
  We are very grateful to S. Hilaire for providing us the HFB gaps
  $\Delta_{n,p}^{(\Gogny)}$.
\end{acknowledgments}
%%%%%%%%%%%%%%%%%%%%%%%%%%%%%%%%%%%%%%%%%%%%%%%%%%%%%%%%%%%%%%%%%%%%%%%%%%%%%%%

\end{document}